\documentstyle[aps,prc,tighten]{revtex}

\begin{document}                             	

\title{ Measuring longitudinal amplitudes for electroproduction of
pseudoscalar mesons using recoil polarization in parallel kinematics }
\author{James J. Kelly}
\address{ Department of Physics, University of Maryland, 
          College Park, MD 20742 }
\date{\today}
\maketitle

\begin{abstract} 
We propose a new method for measuring longitudinal amplitudes for
electroproduction of pseudoscalar mesons that exploits a symmetry
relation for polarization observables in parallel kinematics.
This polarization technique does not require variation of
electron scattering kinematics and avoids the major sources 
of systematic errors in Rosenbluth separation.  
\end{abstract} 

Transition form factors for electroexcitation of nucleon resonances provide
important tests of QCD-inspired models baryon structure.
However, it is often very difficult to separate unpolarized longitudinal
response functions from the dominant transverse response functions using
the traditional Rosenbluth method without substantial systematic errors
arising from the strong dependence of both acceptances and cross sections 
upon electron-scattering kinematics.
Arnold, Carlson, and Gross \cite{Arnold81}
demonstrated that the ratio between electric and magnetic nucleon elastic 
form factors can be measured using either recoil or target polarization; 
such techniques are now becoming standard for elastic scattering.

In this Brief Report we demonstrate that polarization observables for
electroproduction of pseudoscalar mesons in parallel kinematics can be
used to separate longitudinal and transverse amplitudes without need of
Rosenbluth separation.
When the nucleon momentum and spin are both parallel to the momentum
transfer, conditions sometimes described as {\it superparallel kinematics} 
\cite{Boffi93},
the polarized and unpolarized transverse response functions become
identical and the recoil polarization or the target polarization asymmetry
can be used to determine the ratio between longitudinal and transverse
cross sections \cite{Kelly96,TJNAF96-001,TJNAF93-013}.
Some of the implications of this symmetry have been considered for nucleon 
knockout reactions upon spin-0 targets which leave the residual nucleus with
spin-$\frac{1}{2}$ \cite{Boffi88}
and for electron scattering by a polarized spin-$\frac{1}{2}$ target
\cite{Giusti89}.
Raskin and Donnelly \cite{Raskin89} also mention this symmetry for pion
electroproduction.
Using the nonstandard multipole expansion of Raskin and Donnelly and assuming 
dominance of the $M_{1+}$ amplitude for the $N\rightarrow \Delta$ transition, 
Schmieden \cite{Schmieden98} discussed the sensitivity of polarization
observables for pion electroproduction to quadrupole amplitudes,
but the symmetry is more general and it is not necessary to assume that 
a single resonance dominates.
The complete tables of response functions expressed in terms of helicity 
amplitudes for pseudoscalar meson production which can be found in 
Refs.\ \cite{KDT,Drechsel92} implicitly contain the results below also. 
Nevertheless, here we present some of the practical aspects of exploiting
this symmetry because the simplicity and utility of the polarization 
method for superparallel kinematics does not appear to be widely known 
or appreciated. 

The reaction amplitudes for any $A(e,e^\prime N)B$ process where $A$
has spin-$\frac{1}{2}$ and $B$ spin-$0$ that is governed by the
one-photon exchange mechanism can be expressed in terms of helicity 
amplitudes of the form
\begin{equation}
H_{\lambda_f \lambda_i \lambda_\gamma}(Q^2,W,\theta,\phi) =\langle 
\lambda_f | {\cal F}_\mu \varepsilon^\mu | \lambda_i,\lambda_\gamma \rangle 
\end{equation}
where $\lambda_i$ and $\lambda_f$ are the initial and final helicities
of the nucleon, $\lambda_\gamma$ is the helicity of the virtual photon, 
${\cal F}^\mu$ is an appropriately normalized transition current operator,
and $\varepsilon^\mu$ is the virtual-photon polarization vector.
The invariant mass of the meson-nucleon system is given by $W$, 
while $Q^2 = {\bf q}^2 - \omega^2$ is the virtuality of a spacelike photon.
The pion direction relative to the momentum transfer ${\bf q}$ and 
electron-scattering plane is given by polar and azimuthal angles 
$\theta$ and $\phi$.
We label a nucleon recoil momentum that is along ${\bf q}$, such
that $\theta=\pi$, as parallel or a nucleon momentum in
the opposite direction as antiparallel, 
and assign $\phi=0$ to both.   
Phase conventions for helicity states follow the conventions
of Jacob and Wick \cite{Jacob59}.

The present derivation assumes that the reaction is mediated by one-photon 
exchange and conserves parity, but makes no other assumptions about the 
details of the transition amplitudes.  
Since parity conservation \cite{Jacob59} requires
$|H_{-\lambda_f -\lambda_i -\lambda_\gamma}| =
|H_{ \lambda_f  \lambda_i  \lambda_\gamma}| $, 
it is sufficient to consider six independent amplitudes $H_i$ 
for $(\lambda_f,\lambda_i,\lambda_\gamma)$ chosen as 
 $(-\frac{1}{2},-\frac{1}{2}, 1)$, 
 $(-\frac{1}{2}, \frac{1}{2}, 1)$, 
 $( \frac{1}{2},-\frac{1}{2}, 1)$, 
 $( \frac{1}{2}, \frac{1}{2}, 1)$, 
 $( \frac{1}{2}, \frac{1}{2}, 0)$, and 
 $( \frac{1}{2},-\frac{1}{2}, 0)$    
and numbered sequentially \cite{Jones65,Walker69}. 
Due to the absence of orbital angular momentum in the initial
state or spin in the undetected recoil particle ($B$), the angular 
momentum projected onto the virtual photon direction reduces to
$J_z = \lambda_\gamma - \lambda_i = \pm \lambda_f$
for parallel or antiparallel kinematics, where the upper sign applies to 
parallel and the lower to antiparallel kinematics.
Hence, only $H_4$ and $H_6$ contribute to parallel or $H_2$ and $H_5$ to
antiparallel kinematics. 
It is convenient to define $T_+=H_4$ and $L_+=H_6$ as the transverse
and longitudinal amplitudes relevant to parallel kinematics and 
$T_-=-H_2$ and $L_-=H_5$ as the corresponding amplitudes for antiparallel
kinematics.
These amplitudes are related to the usual CGLN \cite{CGLN2,Dennery61}
coefficients by
\begin{mathletters}
\begin{eqnarray}
T_{\pm} &=& \sqrt{2} ({\cal F}_1 \pm {\cal F}_2) \\
L_{\pm} &=& \frac{Q}{\omega}({\cal F}_5^\prime \mp {\cal F}_6^\prime) 
\end{eqnarray}
\end{mathletters}
where
\begin{mathletters}
\begin{eqnarray}
  i {\cal F}^0 &=& \frac{q}{\omega}\left( 
    {\cal F}_5^\prime \vec{\sigma}\cdot\hat{q}
  + {\cal F}_6^\prime \vec{\sigma}\cdot\hat{p} \right) \\
  i \vec{{\cal F}} &=& 
    {\cal F}_1 \vec{\sigma} 
- i {\cal F}_2 \vec{\sigma}\cdot\hat{p} \vec{\sigma}\times\hat{q}
+   {\cal F}_3 \hat{p} \vec{\sigma}\cdot\hat{q}
+   {\cal F}_4 \hat{p} \vec{\sigma}\cdot\hat{p}
+   {\cal F}_5 \hat{q} \vec{\sigma}\cdot\hat{q}
+   {\cal F}_6 \hat{q} \vec{\sigma}\cdot\hat{p}
\end{eqnarray}
\end{mathletters}
and
\begin{mathletters}
\begin{eqnarray}
{\cal F}_5^\prime &=& {\cal F}_5 + {\cal F}_3 \hat{p}\cdot\hat{q} + {\cal F}_1 \\
{\cal F}_6^\prime &=& {\cal F}_6 + {\cal F}_4 \hat{p}\cdot\hat{q} 
\; .
\end{eqnarray}
\end{mathletters}

Using the standard multipole expansion of CGLN amplitudes
introduced by Dennery \cite{Dennery61}, 
the amplitudes for antiparallel kinematics become
\begin{mathletters}
\begin{eqnarray}
T_{-} &=& \sqrt{\frac{1}{2}} \sum_\ell \left[ 
(\ell+1)(\ell+2) E_{\ell +} + \ell (\ell-1) E_{\ell -}
+ \ell (\ell+1)(M_{\ell+} - M_{\ell -}) \right] \\
L_{-} &=& \frac{Q}{q} \sum_\ell \left[ (\ell+1)^2 S_{\ell +} + 
\ell^2 S_{\ell -} \right]
\end{eqnarray}
\end{mathletters}
while the summands for parallel kinematics require an extra factor of 
$(-)^\ell$.
Note that because Raskin and Donnelly \cite{Raskin89} confused 
$\theta_\pi$ with $\theta_N$,
the multipole amplitudes used by Schmieden \cite{Schmieden98} should be 
multiplied by $(-)^\ell$. 
Furthermore, Raskin and Donnelly give the opposite sign for $E_{\ell -}$.

The differential cross section for the meson electroproduction reaction 
$p(\vec{e},e^\prime \vec{N})x$ can be expressed in the form 
\begin{equation}
 \frac{d^5 \sigma}{dk_f d\Omega_e d\Omega_N} =
\Gamma_\gamma \sigma_v
\end{equation}
where $\sigma_v$ is the center of mass cross section for the 
virtual photoproduction reaction  $\gamma_v+N \rightarrow x+N$ and
\begin{equation} 
  \Gamma_\gamma= \frac{\alpha}{2 \pi^2} \frac{k_f}{k_i} 
         \frac{k_\gamma}{Q^2}\frac{1}{1-\epsilon}
\end{equation}
is the virtual photon flux for initial (final) electon momenta $k_i$ ($k_f$).
Here 
$\epsilon = \left( 1+2\frac{{\bf q}^2}{Q^2}\tan^2 \frac{\theta_e}{2} 
\right)^{-1}$ 
is the transverse polarization of the virtual photon,
$\theta_e$ is the electron scattering angle, and 
$k_\gamma = (W^2 - m_p^2)/2m_p$
is the laboratory energy a real photon would need to excite the same 
transition.
The spin dependence of the virtual photoproduction cross section 
for an unpolarized target can be expressed in the form
\begin{equation}
\sigma_v =
   \bar{\sigma} \left[1 + \bbox{P}\cdot \bbox{\sigma} 
   +h (A + \bbox{P}^{\prime} \cdot \bbox{\sigma})\right] 
\end{equation}
where $\bar{\sigma}$ is the unpolarized differential cross section, 
$A$ is the beam analyzing power, 
$\bbox{P}$ is the induced or helicity-independent recoil polarization, 
$\bbox{P}^\prime$ is the polarization transfer or
helicity-dependent recoil polarization, and $h$ is the beam helicity.
Thus, the net polarization of the recoil nucleon is
$\bbox{\Pi} = \bbox{P} + h\bbox{P}^\prime$.
A similar expression applies when the target is polarized and the
recoil polarization is unobserved.
We omit observables requiring both target and recoil polarization 
because they provide no new information for parallel kinematics 
and are so difficult to measure as to be of no practical interest.
For parallel kinematics it is simplest to refer polarizations to a basis 
in which $\bbox{\hat{z}}=\bbox{\hat{q}}$ is in the photon direction,
$\bbox{\hat{y}} = \bbox{\hat{k}}_i \times \bbox{\hat{k}}_f$ is normal to 
the electron scattering plane, and 
$\bbox{\hat{x}}=\bbox{\hat{y}}\times\bbox{\hat{z}}$ is transverse.

Recoil polarization observables can now be expressed in the form
\begin{mathletters}
\begin{eqnarray}
\bar{\sigma} &=& \sigma_{T} + \epsilon \sigma_{L} 
= {\cal K} ( \frac{1}{2}|T|^2 + \epsilon |L|^2 ) \\
\Pi_x \bar{\sigma} &=&  h {\cal K} 
\sqrt{\epsilon (1-\epsilon)} Re(T L^\ast) \\
\Pi_y \bar{\sigma} &=& - {\cal K} 
\sqrt{\epsilon (1+\epsilon)} Im(T L^\ast) \\
\Pi_z \bar{\sigma} &=& h {\cal K} \sqrt{1-\epsilon^2} \frac{1}{2} |T|^2 
\end{eqnarray}
\end{mathletters}
where ${\cal K} =  p W / k_\gamma m_p$ is a kinematic factor and $p$ is
the final center of mass momentum.
We have left the $\pm$ subscripts on observables and amplitudes implicit
in the interests of brevity.
In the chosen basis target and recoil polarization observables for parallel
kinematics differ only in sign.
Thus, using either recoil or target polarization, there are five
observables that depend upon just four response functions 
(bilinear amplitude products).
Therefore, there exists a relationship between polarization and cross section 
for parallel kinematics that provides an alternative method for separating 
the longitudinal and transverse cross sections.

We define 
\begin{equation}
{\cal R}_{\pm} = \frac{\sigma_{L\pm}}{\sigma_{T\pm}} 
= 2\frac{|L_\pm|^2}{|T_\pm|^2}
\end{equation} 
as the ratio between longitudinal and transverse cross sections for
parallel kinematics.
The traditional Rosenbluth separation method relies on the variation of
cross section with $\epsilon$, 
but this method requires measurements for two or more electron scattering
kinematics with quite different acceptances.
When the longitudinal contribution is small, the systematic errors due
to acceptances and kinematic variables can become prohibitively large.
Alternatively, for parallel kinematics it is possible to exploit the
relationship between the longitudinal component of 
recoil (or target) polarization and the transverse contribution to the
differential cross section to obtain
\begin{equation}
{\cal R} = \frac{h \sqrt{1-\epsilon^2} - \Pi_z}{\epsilon \Pi_z}
\end{equation}
with fixed electron scattering kinematics without Rosenbluth separation;
in fact, for the ratio one does not even need to normalize the cross section.
Therefore, this polarization technique avoids the major sources of
systematic error that afflict the Rosenbluth method.

Assuming that $\epsilon$ is known accurately, the uncertainty in ${\cal R}$
is related to the polarization uncertainty $\delta \Pi_z$ by
\begin{equation}
\frac{\delta{\cal R}}{\delta\Pi_z} = 
\frac{|h|\sqrt{1-\epsilon^2}}{\epsilon \Pi_z^2} = 
\frac{(1+{\cal R}\epsilon)^2}{|h|\epsilon\sqrt{1-\epsilon^2}} \; .
\end{equation}
If ${\cal R}$ is small and if the minimum attainable value of $\delta\Pi_z$
is governed by systematic errors, then the optimum kinematics for 
measurement of ${\cal R}$ are realized when $\epsilon =1/\sqrt{2}$.
Alternatively, when statistical uncertainties dominate $\delta\Pi_z$,
it becomes advantageous to employ the largest practical $\epsilon$
because the virtual-photon flux is proportional to $(1-\epsilon)^{-1}$;
for given $W$ and $Q^2$ this implies that higher beam energies are
more favorable.

These relationships can also be used to establish a bound 
\begin{equation}
|\Pi_z| \leq |h| \sqrt{1-\epsilon^2}
\end{equation}
upon the longitudinal polarization, 
where the limiting value is realized for a purely transverse 
electroproduction amplitude.
Recognizing that in parallel kinematics the transverse amplitudes flip
nucleon spin while the longitudinal amplitudes do not, 
one finds that reduction of the longitudinal polarization from its
maximal value would be indicative of a measurable spin nonflip amplitude.
Furthermore, a nonvanishing normal component of polarization is indicative
of a phase difference between longitudinal and transverse amplitudes.
The magnitudes of the longitudinal and transverse helicity amplitudes
and the relative phase between them can be determined from
\begin{mathletters}
\begin{eqnarray}
L &=& r T e^{i\delta} \\
|T|^2 &=& \frac{2 \bar{\sigma} \Pi_z}{h {\cal K} \sqrt{1-\epsilon^2}} \\
r^2 &=& {\cal R}/2 = \frac{h \sqrt{1-\epsilon^2} - \Pi_z}{2 \epsilon \Pi_z} \\
\tan{\delta} &=& \sqrt{\frac{1-\epsilon}{1+\epsilon}}
\frac{\Pi_y}{\Pi_x} h  \; .
\end{eqnarray}
\end{mathletters}

Finally, under some conditions these quantities provide useful constraints
upon multipole amplitudes.
For example, if we limit the expansions to $s$- and $p$-waves and include 
only contributions involving the dominant $M_{1+}$ amplitude for
pion electroproduction near the $P_{33}(1232)$ resonance, we find
\begin{equation}
Re\left(S_{1-} + 4 S_{1+} \mp S_{0+}  \right)M_{1+}^\ast \approx 
\frac{\Pi_{x\pm} \bar{\sigma}_{\pm}}{ h \sqrt{2}\frac{Q}{q}{\cal K} 
\sqrt{\epsilon-\epsilon^2} }
 \; .
\end{equation} 
Thus, by comparing parallel versus antiparallel kinematics one can separate
the combinations $Re S_{0+}M_{1+}^\ast$ and $Re(S_{1-} + 4 S_{1+})M_{1+}^\ast$.
Most attempts to measure the $S_{1+}$ amplitude for pion electroproduction, 
which is sensitive to quadrupole deformation of the nucleon and $\Delta$ wave 
functions, have relied upon the $R_{LT}$ response function obtained from 
the left-right asymmetry of the unpolarized cross section.
However, Mertz {\it et al.} \cite{Mertz99} have shown that current models fail
to reproduce the $W$ dependence of the cross section asymmetry, which casts
doubt upon the reliability of fitted $S_{1+}$ resonance amplitudes.
It is widely believed that the $S_{0+}$ contribution may be responsible for 
these difficulties, 
but it should be possible to measure this amplitude with relatively
little model dependence using recoil polarization for parallel kinematics.
Furthermore, although one cannot separate $S_{1+}$ from $S_{1-}$ without more 
comprehensive data, the $S_{1-}$ contribution is expected to be quite small 
and to display a distinctly different dependence upon $W$.
Therefore, recoil polarization for parallel versus antiparallel kinematics 
offers an independent method for measuring $S_{1+}$ also.

In summary, we have proposed a polarization method for measuring the ratio
between longitudinal and transverse cross sections for electroproduction of
pseudoscalar mesons in parallel kinematics that employs fixed electron
kinematics and avoids Rosenbluth separation.
We have also developed the relationships needed to extract the
magnitudes and relative phase of the corresponding helicity amplitudes.
It is important to recognize that this method does not depend upon
dominance of any particular resonance and applies equally well to the 
resonant and nonresonant contributions.
However, it need not apply to more complicated background processes
such as $p(e,e^\prime N) \pi \pi$.
Fortunately, for many interesting experiments, such as 
$\gamma_v N\rightarrow P_{33}(1232) \rightarrow N\pi$ or
$\gamma_v N\rightarrow S_{11}(1535) \rightarrow N\eta$, 
those background contributions should vary slowly with missing mass and can 
be subtracted from the single-meson peak in the missing mass distribution.
Nor can this method be used to obtain full angular distributions 
for longitudinal and transverse response functions.
Nevertheless, the ability to separate longitudinal and transverse
amplitudes in parallel and/or antiparallel kinematics using recoil 
polarization measurements without Rosenbluth separation can be 
very helpful in testing models of baryon structure and is a useful 
supplement to the traditional cross section method.


\end{document}